# A Spiking Neural Network Model of Elementary Self-Consciousness via Endogenous Default Mode Network Dynamics

R. Lahoz-Beltra

Modeling, Data Analysis and Computational Tools for Biology Research Group, Complutense University of Madrid, 28040 Madrid, Spain

lahozraf@ucm.es

The Crazy Lab©, Madrid, Spain.

rafa.lahoz@protonmail.ch

**Abstract**

Understanding the neurobiological mechanisms underlying self-referential cognition and baseline self-consciousness remains a fundamental challenge in computational neuroscience. In this work, we propose a large-scale computational model incorporating a 10,000-neuron spiking neural network (SNN) based on Izhikevich dynamics. The network is structured into two interacting subsystems: a sensory processing layer (5,000 regular-spiking cortical neurons) and an endogenous Default Mode Network (DMN) pacemaker subsystem (5,000 intrinsically bursting neurons). The DMN layer is modulated by continuous tonic currents reflecting ascending brainstem neuromodulation, maintaining intrinsic, autonomous bioelectric rhythms independent of external sensory input. To represent top-down cognitive modulation, synaptic weights are hierarchically structured such that DMN-to-network projections exceed sensory-level connections. Through numerical simulations using a modified two-step Euler integration scheme, we demonstrate how endogenous pacemaker activity interacts with transient external sensory perturbations, providing an elementary mathematical framework for the emergence of a persistent, self-sustaining neural representation of “Self”.

## 1. Introduction

Self-consciousness—the subjective experience of an enduring “Self” capable of autonomous mental activity—is deeply rooted in endogenous brain dynamics. Neuroimaging studies (Raichle et al., 2001; Raichle, 2015) have consistently associated self-referential processing, mind-wandering, and baseline self-awareness with the Default Mode Network (DMN), a distributed anatomical network comprising the posterior cingulate cortex, medial prefrontal cortex, and thalamic (Steriade, 2006) hubs. Unlike strictly sensory or motor-driven cortical networks, the DMN maintains intrinsic, rhythmic bioelectric activity even in the absence of goal-directed tasks or sensory inputs.

Computational models of conscious processing often focus on feedforward sensory hierarchies or global workspace architectures. However, capturing the transition from passive sensory reception to active, self-sustaining internal states requires biologically realistic spiking neural networks (SNNs) capable of exhibiting rich dynamical behaviors, such as intrinsic bursting and pacemaker oscillations.

In this paper, we formulate an elementary computational model of baseline self-consciousness (Figure 1). Using Izhikevich's two-variable differential framework (Izhikevich, 2003), we construct a 10,000-neuron network divided into a sensory layer and a DMN pacemaker module. We demonstrate how tonic neuromodulatory currents and asymmetrical top-down synaptic weights maintain an autonomous bioelectric "pulse of the Self" amidst incoming sensory perturbations.

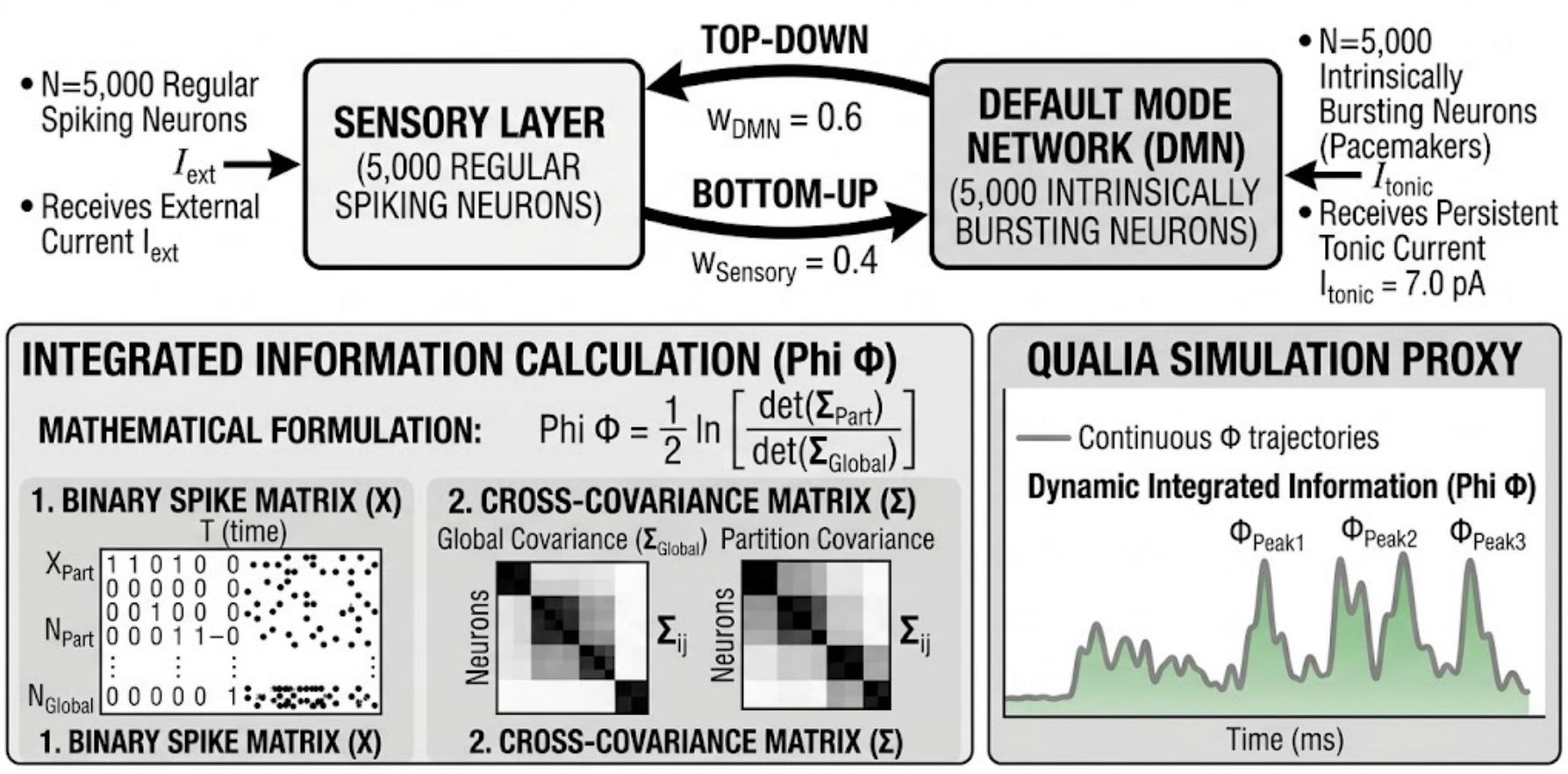

**Figure 1. Architecture and quantitative framework of the 10,000-neuron Spiking Neural Network (SNN) for computational consciousness and Qualia simulation.** Top: Network architecture comprising a Sensory Layer ($N$=5,000 regular spiking neurons) driven by external current $I_{ext}$, bidirectionally coupled via top-down ($w_{DMN}$ = 0.6) and bottom-up ($w_{Sensory}$ = 0.4) weights to a Default Mode Network (DMN; $N$=5,000 intrinsically bursting pacemaker neurons) subject to persistent tonic current $I_{tonic}$ = 7.0 pA. Bottom left: Mathematical formulation and matrix transformations for computing integrated information ($\phi$), utilizing binary spike matrices ($X$) and global vs. partition cross-covariance matrices ($\sum_{Global}$, $\sum_{Part}$). Bottom right: Continuous dynamic $\phi$ trajectories reflecting simulated Qualia states over time.

## 2. Model Formulation and Biophysical Dynamics

### 2.1 Bioelectric Engine: Izhikevich Spiking Dynamics

The biophysical behavior of individual neurons in the network is governed by the two-dimensional system of coupled ordinary differential equations proposed by Izhikevich (2003):

$$\frac{dv}{dt} = 0.04v^2 + 5v + 140 - u + I \quad (1)$$

$$\frac{du}{dt} = a(bv - u) \quad (2)$$

with the auxiliary after-spike resetting rule:

$$\text{if} \quad v \geq 30\,mV \text{ then} \begin{cases} v \leftarrow c \\ u \leftarrow u + d \end{cases} \quad (3)$$

In the model, $v$ represents the membrane potential, $u$ (pA) denotes the membrane recovery variable (providing negative feedback to $v$), and $I$ (pA) is the total input current combining external stimuli, synaptic integration, and intrinsic tonic drive. The dimensionless parameters $a$, $b$, $c$, and $d$ define the dynamic firing regimes: $a$ describes the time scale of the recovery variable $u$; $b$ specifies the sensitivity of $u$ to subthreshold fluctuations of $v$; $c$ represents the post-spike reset value of $v$; and $d$ describes the post-spike reset increment of $u$.

2.2 Network Architecture and Heterogeneity

The network comprises $N$ = 10,000 spiking neurons divided into two functional sub-networks (Figure 2):

Sensory Processing Layer ($i$ = $0\ldots4{,}999$).- Models cortical excitatory pyramidal neurons exhibiting regular spiking (RS) dynamics. To capture physiological heterogeneity across the cortical population, individual resetting parameters $c_i$ and $d_i$ are randomly drawn from a uniform distribution:

$$c_i = -65.0 + 15.0\,r_i^2, \quad d_i = 8.0 - 6.0\,r_i^2 \quad (4)$$

where $r_i \sim U(0, 1)$. The baseline reset potential centered around -65mV reproduces typical cortical excitatory profiles.

Default Mode Network (DMN) Pacemaker Subsystem ($i$ = $5{,}000\ldots9{,}999$).- Models pacemaker-like bursting units representative of thalamocortical and posterior cingulate assemblies. Within this pool, a designated core of pacemaker neurons ($i$ = $5{,}000\ldots\ 5{,}999$) is assigned deterministic parameter values, i.e., $c_i = -55.0$ mV and $d_i = 4.0$. Elevating $c_i$ to -55 mV shortens the refractory period, inducing intrinsically bursting (IB) behavior. This generates an autonomous bioelectric pulse representing the basal core of the “Self”.

2.3 Synaptic Connectivity and Hierarchical Topology

The total network exhibits a sparse connectivity structure with $K$ = 1,000 random outgoing projections per neuron, yielding $10^7$ total synaptic connections stored in a sparse format. The synaptic weight matrix $S_{ij}$ between presynaptic neuron $j$ and postsynaptic neuron $i$ is defined by:

$$S_{ij} = U(0,1)\ w_{ij} \quad (5)$$

where $w_{ij}$ follows a top-down hierarchical decision function (Figure 1):

$$w_{ij} = \begin{cases} 0.6 & j \geq 5000 \text{( DMN projections )} \\ 0.4 & j < 5000 \text{( Sensory projections )} \end{cases} \quad (6)$$

This asymmetry reflects the functional supremacy of the DMN over sensory cortical areas (Varela et al., 2001). Top-down modulatory projections from self-referential assemblies exert stronger influence ($w$ = 0.6) over lower-level sensory regions ($w$ = 0.4), enabling the “Self” network to effectively modulate sensory representations.

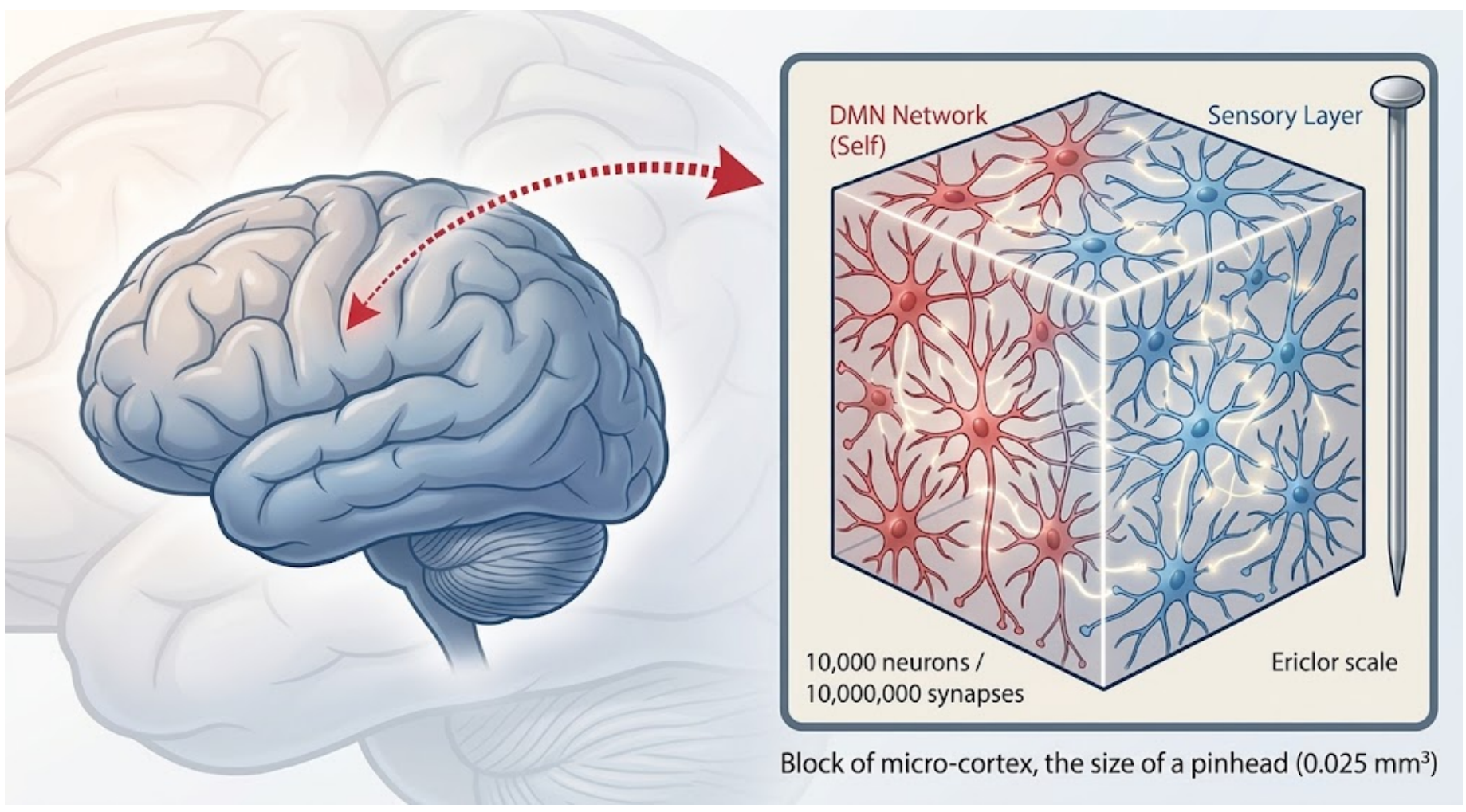

**Figure 2**. A 0.025 $mm^3$ micro-cortex block, equivalent to a pinhead, containing 10,000 neurons and 10 million synapses, showing a DMN-sensory layer boundary.

## 2.4 Biological Model of Self-Consciousness and Input Currents

In every simulation millisecond, the total current $I_i(t)$ arriving at neuron $i$ integrates endogenous, exogenous, synaptic, and stochastic inputs (Figure 3):

$$I_i(t) = I_{i,tonic} + I_{i,ext} + I_{i,syn} + \eta_i(t) \quad (7)$$

Endogenous Tonic Current ($I_{i,tonic}$).- Represents ascending dopaminergic and cholinergic brainstem neuromodulation critical for maintaining wakefulness and intrinsic DMN activity. It is continuously applied to the core DMN pacemaker neurons ($i$ = 5,000… 5,999):

$$I_{i,tonic} = \begin{cases} 7.0 \text{ pA if } 5000 \leq i \leq 5999 \\ 0.0 \quad \text{otherwise} \end{cases} \quad (8)$$

Exogenous Transient Stimuli ($I_{i,ext}$).- External sensory input is modeled as a step-current applied to a subpopulation of 1,500 sensory neurons during a defined temporal window (300 ms< $t$ < 600 ms):

$$I_{i,ext} = \begin{cases} 16.0 \text{ pA} & \text{if } 0 \le i < 1500 \\ 0.0 & \text{otherwise} \end{cases} \quad (9)$$

Synaptic Current Integration ($I_{i,syn}$).- Propagates action potentials across the connectivity matrix. The integrated synaptic current $I_{i,syn}$ models the transmission of a signal when a presynaptic neuron *j* fires a spike at time *t*, propagating the current to all postsynaptic neurons *i*. Here, *j* denotes the set of presynaptic neurons that have fired within that specific millisecond, i.e., $\{ j \mid v_j(t) \ge 30 \text{ mV} \}$. Consequently, this incoming synaptic current alters the internal synchrony of the "Self" driving a transition from an internal baseline state to an attentional state:

$$I_{i,syn}(t) = \sum_j S_{ij} \quad (10)$$

Cognitive Noise ($\eta_i(t)$).- Unstructured background fluctuations, i.e., thermal and stochastic brain noise, is modeled as uncorrelated Gaussian $N(0,\sigma^2)$ noise with $\sigma = 3$ in the simulation experiment conducted in this study.

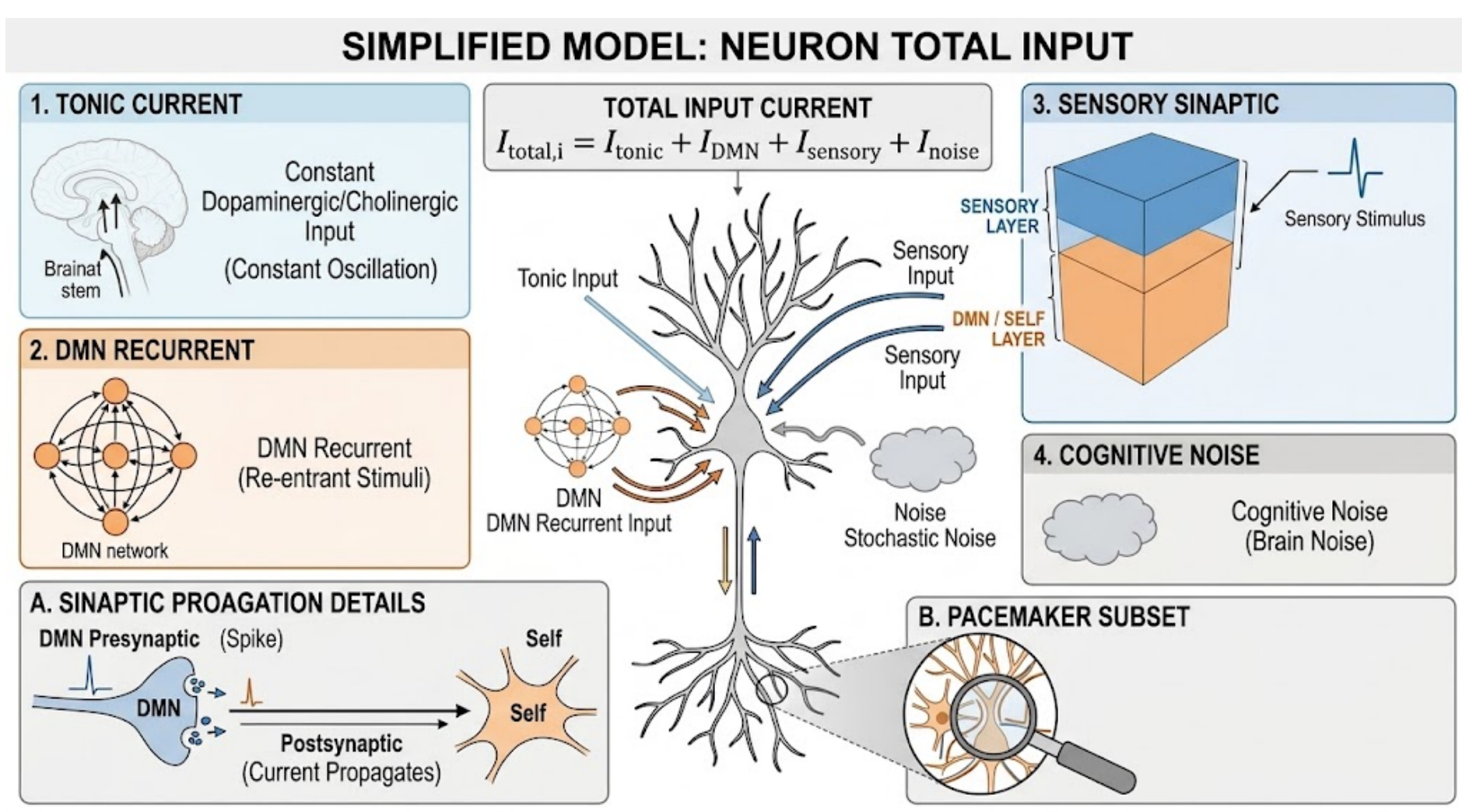

**Figure 3.** Simplified model of total neuronal input current. The total current $I_{total,i}$ integrates four distinct sources: (1) tonic input ($I_{tonic}$) from neuromodulatory systems, (2) recurrent DMN input ($I_{DMN}$) representing re-entrant stimuli, (3) sensory synaptic input ($I_{sensory}$), and (4) stochastic cognitive noise ($I_{noise}$). Insets depict (A) presynaptic to postsynaptic spike propagation and (B) the pacemaker subset within the neural architecture.

## 2.5 Numerical Integration

To balance numerical stability and computational efficiency while preventing the overhead of 4th-order Runge-Kutta schemes, the network dynamics are evaluated using a modified two-step (half-step) Euler integration method with a fixed time step $\Delta t = 1.0$ ms:

$$v_i^1 = v_i(t) + 0.5\left(0.04v_i(t)^2 + 5v_i(t) + 140 + u_i(t) + I_i(t)\right) \quad (11)$$

$$v_i(t+1) = v_i^1 + 0.5\left(0.04\left(v_i^1\right)^2 + 5v_i^1 + 140 + u_i(t) + I_i(t)\right) \quad (12)$$

$$u_i(t+1) = u_i(t) + a_i\left(b_i v_i(t+1) - u_i(t)\right) \quad (13)$$

2.6 Quantification of Conscious Experience and Qualia via Integrated Information Theory (IIT)

To quantify the emergence and richness of conscious experience (Qualia), we adapt the principles of Integrated Information Theory (IIT) (Tononi, 2004; Balduzzi and Tononi, 2008, 2009). IIT posits that consciousness corresponds to the capacity of a system to integrate non-redundant information—meaning that the information generated by the unified whole transcends the sum of information generated independently by its isolated parts (Figure 4). The scalar metric $\phi$ quantifies this degree of informational integration and characterizes the information content of subjective experience. Within our spiking network model, $\phi$ measures the degree of dynamic functional coupling between the sensory layer and the Default Mode Network (DMN), capturing how external sensory perturbations modify and covary with the endogenous “Self” dynamics.

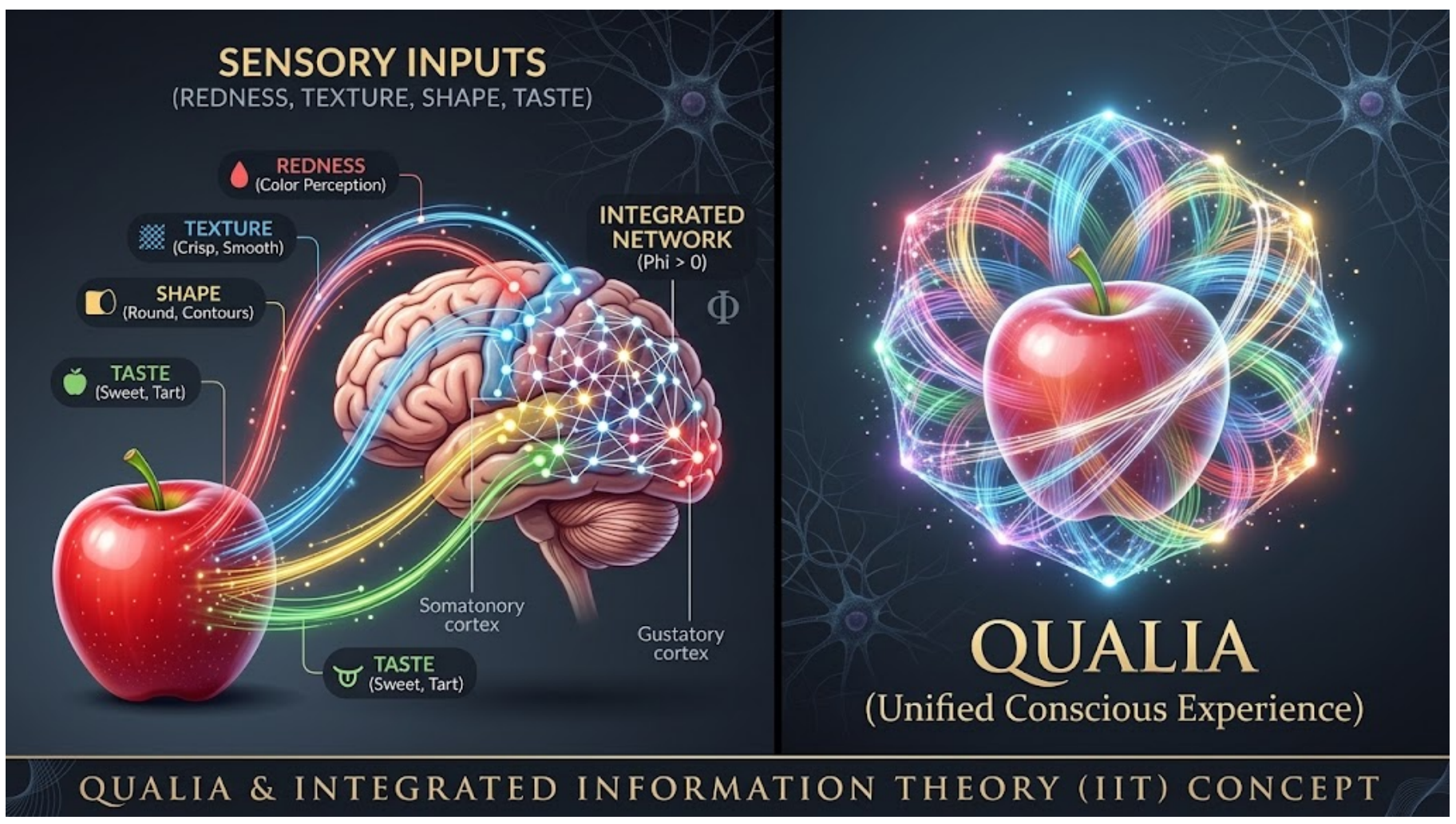

**Figure 4.** Conceptual illustration of Qualia and Integrated Information Theory (IIT). Multi-modal sensory inputs (redness, texture, shape, and taste) are processed across distinct cortical regions and integrated into a high-$\phi$ network ($\phi > 0$). This integration transforms separate sensory features of an apple into a single, unified subjective conscious experience (Qualia)—the subjective, indivisible “what it is like” to perceive the apple as a whole.

*2.6.1 Temporal Window Selection and Binary Spiking Matrix*

The estimation of proceeds across discrete analysis windows of duration $\tau$ ms (e.g., $\tau$ = 5 ms, $\tau$ = 80 ms). Within each evaluation window, we isolate the subset of $M$ most active spiking neurons (with $M \leq 200$) across the network. We define a binary state matrix $\{0, 1\}^{M \times \tau}$, where the element $X_{i,t}$ indicates whether neuron $i$ fired an action potential at time $t$:

$$X_{i,t} = \begin{cases} 1 & \text{if neuron } i \text{ spiks at time } t \\ 0 & \text{otherwise} \end{cases} \quad (14)$$

For instance, if $M$ active neurons are selected, $X_{1,1} = 1$ denotes that neuron 1 fired during the first millisecond of the observation window, $X_{2,2} = 1$ denotes a spike by neuron 2 in the second millisecond, and so on.

*2.6.2 Firing Rates and Cross-Covariance Matrix*

The mean firing rate $\mu_i$ for each selected neuron $i$ across the time window $\tau$ is computed as:

$$\mu_i = \frac{\sum_{t=1}^{\tau} X_{i,t}}{\tau} \quad (15)$$

To evaluate the degree of fine-scale dynamic synchrony across the population, we calculate the pairwise cross-covariance $\sum_{ij}$ between neurons $i$ and $j$:

$$\sum_{ij} = \frac{1}{\tau} \sum_{t=1}^{\tau} \left( X_{i,t} - \mu_i \right) \left( X_{j,t} - \mu_j \right) \quad (16)$$

A positive covariance $\sum_{ij} > 0$ indicates systematic co-activation or phase-locking between neurons $i$ and $j$ within the temporal window $\tau$ (Barrett and Seth, 2011; Mediano et al., 2019). The complete population interaction is thus represented by the global empirical covariance matrix $\sum_{Global} \in \mathbb{R}^{M \times M}$.

*2.6.3 System Bipartition and Integrated Information ($\phi$)*

To assess whether the joint sensory–DMN system generates unified, subjective and irreducible conscious experience (Qualia), the $M$ active neurons matrix is partitioned according to their anatomical sub-networks into two disjoint smaller submatrices: the submatrix $\sum_{Sensory}$ containing active sensory cortical neurons, and submatrix $\sum_{DMN}$ containing active DMN pacemaker neurons. From this bipartition, we construct the partitioned block-diagonal covariance matrix $\sum_{Part}$, which treats the two subsystems as functionally independent:

$$\sum_{Part} = \begin{pmatrix} \sum_{Sensory} & 0 \\ 0 & \sum_{DMN} \end{pmatrix} \quad (17)$$

where $\sum_{Sensory}$ and $\sum_{DMN}$ are the intra-system covariance matrices evaluated independently for the sensory and DMN subpopulations.

Following, Gaussian and covariance-based approximations of integrated information for continuous/spiking dynamics (Barrett & Seth, 2011; Oizumi et al., 2016; Mediano et al., 2019), i.e., the overall integrated information $\phi$, is quantified by evaluating the matrix determinant ratio (or log-likelihood divergence) between the full system matrix $\sum_{Global}$ and its partitioned counterpart $\sum_{Part}$ :

$$\phi = \frac{1}{2}\ln\left(\frac{\det(\sum_{Part})}{\det(\sum_{Global})}\right) \quad (18)$$

When cross-network correlations between sensory and DMN neurons are zero then $\det(\sum_{Global}) = \det(\sum_{Part})$, yielding to $\phi$=0 indicating a complete functional independence and an absence of integrated subjective experience. Conversely, strong reciprocal covariance induced when sensory inputs alter the baseline DMN dynamics drives to $\det(\sum_{Global}) < \det(\sum_{Part})$, increasing $\phi$ and signaling the emergence of an integrated phenomenal of consciousness or Qualia state.

## 3. Results

To evaluate how endogenous pacemaker activity interacts with sensory inputs to produce integrated experience, we simulated the 10,000-neuron spiking network over a 900 ms trajectory. Implementation of the simulation model was carried out using Python 3.13 on an iMac (Apple M4 chip, 16 GB RAM) under the macOS Tahoe 26.3.1 operating system. Figure 5 displays the results of the experiment. The top panel illustrates the raster plot separating the Sensory Layer ($i$ = 0… 4,999, blue) and the DMN Network ($i$ = 5,000 …9,999, red). The middle panel displays the mean population firing rate across the network, and the bottom panel tracks the continuous evolution of integrated information ($\phi$), serving as the proxy for phenomenal Qualia.

The network dynamic exhibits distinct transitions between localized baseline activity and massive, highly synchronized population bursts. Endogenous basal activity is observed during quiescent windows, e.g., $t \in [150, 300]$ ms, $t \in [430, 480]$ ms, and $t \in [650, 780]$ ms, remaining silent the sensory population, while a subset of core DMN pacemaker units ($i$ = 5,000…6,000) maintains persistent, asynchronous low-frequency firing driven by the 7.0 pA tonic current ($I_{tonic}$). This reflects the baseline bioelectric "pulse of the Self". Furthermore, four major synchronized population events occur across the 900 ms window—centered around $t \approx 30$ ms, $t \approx 330$ ms, $t \approx 550$ ms, and $t \approx 830$ ms.

During these transient periods, strong recurrent excitation within the DMN and top-down modulatory projections ($w_{DMN}$ = 0.6) recruit the sensory layer, driving global population firing rates to saturation ($\approx$1000 Hz).

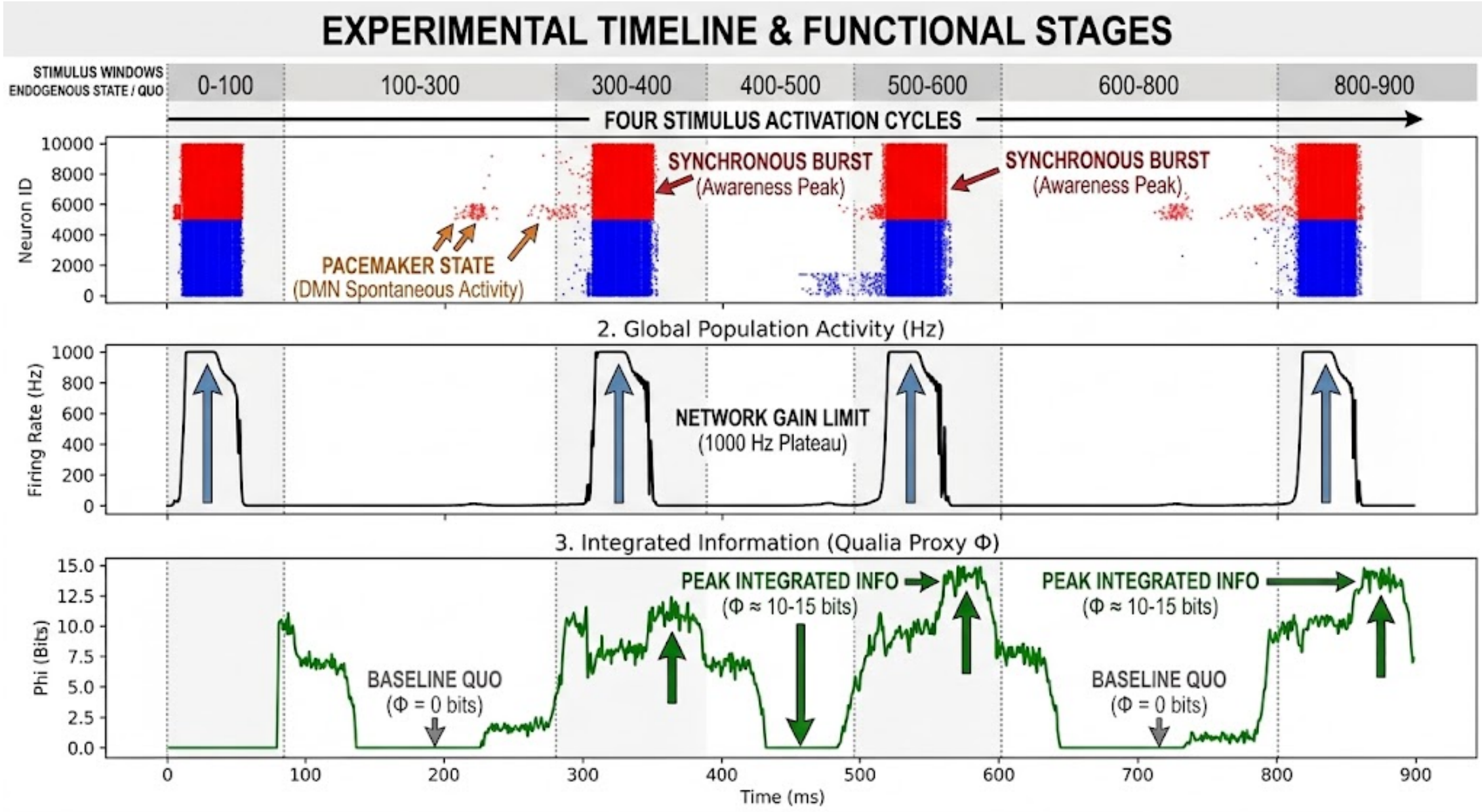

**Figure 5.** Spatiotemporal dynamics and integrated information ($\phi$) of the spiking neural network model. (Top) Raster plot illustrating the spiking activity across $N$ = 10,000 neurons, separated into the Sensory Layer ($i$ = 0...4,999; blue) and the Default Mode Network (DMN; $i$ = 5,000...9,999; red). Persistent basal activity in core DMN pacemakers alternates with highly synchronized population bursts across both subpopulations. (Middle) Global population firing rate (Hz) showing transition dynamics between quiescent low-firing regimes (0-10 Hz) and saturated burst states (1000 Hz). (Bottom) Continuous trajectory of Integrated Information ($\phi$, measured in bits), serving as a quantitative proxy for phenomenal Qualia. Peak values of (12.5-15.0 bits) occur during phases of strong cross-network functional coupling and mutual co-activation between sensory and DMN units.

As shown in Figure 5, the temporal trajectory of $\phi$ reveals a tight non-linear dependence on both the degree of cross-subnetwork co-activation and firing rate synchronization. Three distinct states can be observed: Quiescent or isolated states ($\phi \approx 0$ bits), pre-burst priming states ($\phi \approx$1-2.5 bits) and peak integrated states ($\phi \approx$10-15 bits).

Isolated states with $\phi$ dropping to zero are observed in the absence of sensory-DMN interactions (such as during the prolonged rest state from 150 ms to 220 ms). Here, because sensory networks are inactive and DMN pacemakers fire independently, the global covariance matrix degenerates to a block-diagonal form $\sum_{Global} \approx \sum_{Part}$, yielding to negligible integrated information.

Immediately preceding major population bursts (e.g., around $t \approx$230-280 ms and $t \approx$ 730-780 ms), pre-burst priming states are present and characterized by sparse presynaptic spiking, begining to bridge the sensory layer and the DMN pacemakers. This subtle functional coupling breaks independent block structures, causing $\phi$ to ramp up prior to full global ignition.

Following the previous stages, a third phase emerges, characterized by peak integrated states. As large-scale population bursts ignite, cross-covariances between sensory and DMN neurons surge. The highest peaks in integrated information ($\phi \approx$12.5-15.0 bits) occur during the sustained co-activation phase of the bursts (notably during the third and fourth event windows at $t \approx$ 530-590 ms and $t \approx$820-880 ms). During these windows, the mutual information shared between the DMN ("Self") and sensory populations reaches maximum capacity, demonstrating the formal emergence of an integrated phenomenal state of consciousness or Qualia.

To demonstrate that the observed integrated information dynamics ($\phi$) are structurally robust and independent of specific random network initializations, we expanded our experimental pipeline to a multi-seed benchmark ($N$ = 30) independent realizations. Furthermore, to evaluate the causal role of brainstem-driven tonic neuromodulation, we performed a systematic ablation experiment comparing the control condition ($I_{\text{tonic}}$ = 7.0 pA) against an ablation condition ($I_{\text{tonic}}$ = 0.0 pA), where DMN pacemaker drive is completely suppressed (Figure 6).

Statistical testing via paired $t$-tests ($N$ = 30) confirmed highly significant differences in $\phi(t)$ trajectories across three key functional windows:

1. Baseline Resting State (0 - 300 ms): In the control condition, persistent tonic drive maintained an integrated information baseline of $7.87 \pm 0.12$ bits. Upon ablation ($I_{\text{tonic}}$ = 0.0 pA), baseline integration dropped significantly to $4.67 \pm 0.57$ bits, yielding a $t$-statistic of 30.0614 and $p$ = 2.0750 x $10^{-23}$.
2. Stimulus Onset Latency (300 - 450 ms): Following external sensory input ($I_{\text{ext}}$ = 16.0 pA applied to 1,500 sensory neurons), the control network transitioned rapidly into a high-$\phi$ integrated state ($14.27 \pm 0.76$ bits), whereas the ablated network exhibited a severely delayed and attenuated response ($6.42 \pm 0.10$ bits), yielding a substantial separation, resulting in a $t$-value of 55.9764 with $p$ = 4.2023 x $10^{-31}$.
3. Post-Stimulus Recovery (600 - 900 ms): After stimulus offset, the ablated network rapidly collapsed to zero integration ($\phi \rightarrow 0$ bits), averaging $5.23 \pm 0.40$ bits in the overall window. In contrast, the control network sustained strong baseline re-entry dynamics ($13.13 \pm 1.35$ bits), confirming the necessity of tonic drive for post-perturbation recovery, revealing a statistically significant difference, $t$-value of 32.3387 and $p$ = 2.6493 x $10^{-24}$.

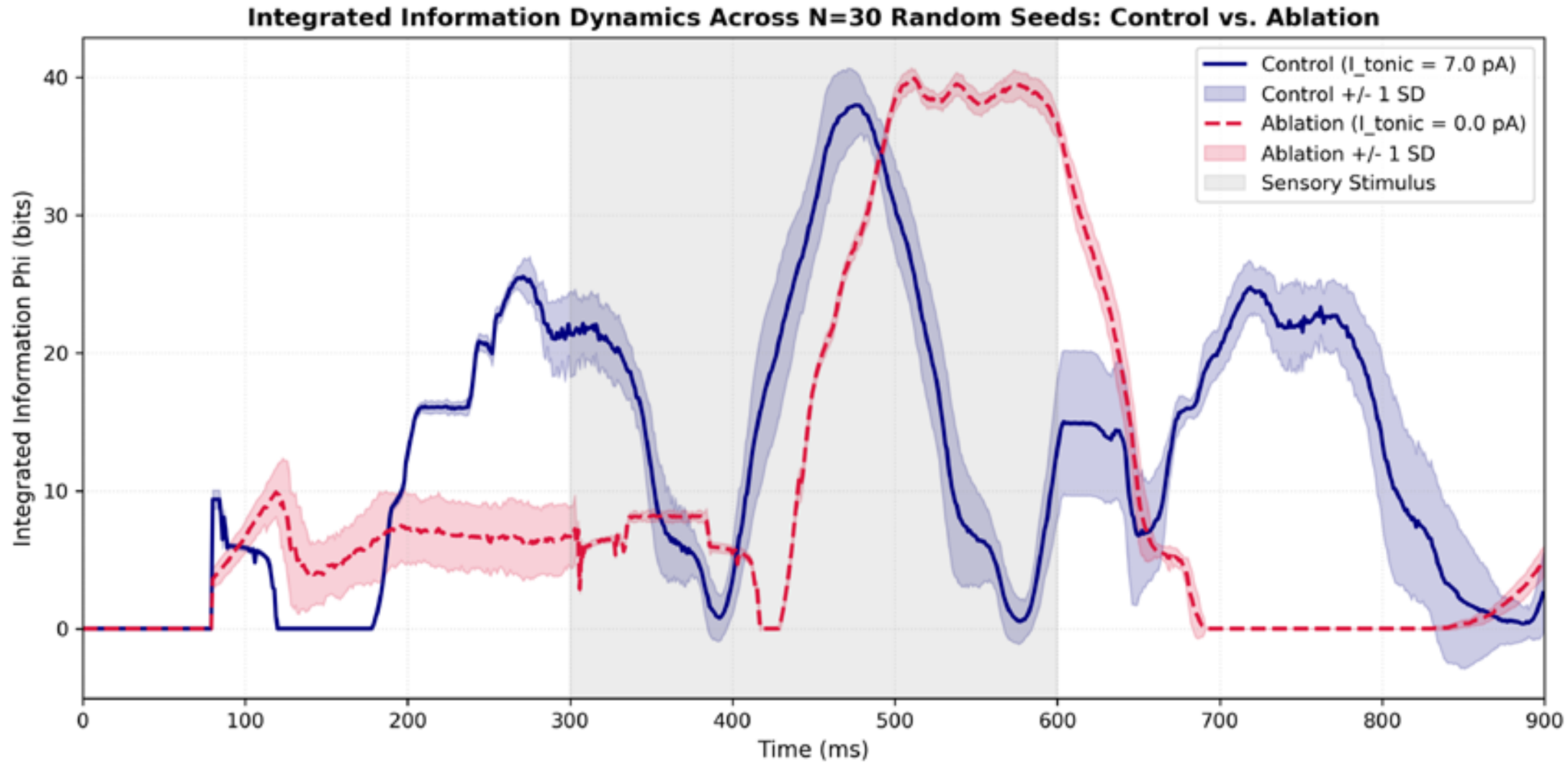

**Figure 6.** Statistical robustness and ablation analysis across $N$ = 30 random initializations. Dynamic trajectories of Integrated Information ($\phi$, mean $\pm$ 1 SD) for the control condition ($I_{tonic}$ = 7.0 pA, blue solid line) versus the ablation condition ($I_{tonic}$ = 0.0 pA, red dashed line). Gray shaded region indicates the external sensory stimulus window (300 - 600 ms). Shaded boundaries represent standard deviations across 30 independent random network topologies, demonstrating robust statistical reproducibility ($p < 10^{-22}$ across all testing windows).

## 4. Discussion

In this work, we formulated a biophysically grounded spiking neural network model that integrates Izhikevich neuronal dynamics with principles from Integrated Information Theory (IIT). By structuring the network into a Regular Spiking sensory layer ($N$ = 5,000) and an intrinsically bursting Default Mode Network (DMN) pacemaker layer ($N$ = 5,000), we reproduced two essential features of baseline cognitive architecture: an endogenous bioelectric "pulse of the Self" driven by persistent tonic currents, and asymmetrical top-down control ($w_{DMN}$ = 0.6) over sensory processing ($w_{Sensory}$= 0.4). Our quantitative evaluation of integrated information ($\phi$) demonstrates that during periods of isolated baseline activity, when sensory neurons remain silent and DMN pacemakers fire asynchronously, $\phi$ remains near zero. The functional covariance matrix during these quiescent phases degenerates into independent block structures, indicating that endogenous activity alone—without cross-network interaction—does not yield unified information. Conversely, maximum values of $\phi$ ($\approx$12.5-15.0 bits) emerge during transient population bursts when incoming sensory perturbations trigger widespread co-activation and phase synchrony across the sensory–DMN boundary. This supports the hypothesis that integrated phenomenal states (Qualia) emerge not from static architectural connectivity alone, but from the dynamic, non-linear coupling between internal self-referential dynamics and sensory inputs (Varela et al., 2001; Tononi, 2004; Deco et al., 2011).

The multi-seed benchmark ($N$ = 30) firmly establishes that the dynamic emergence of high-$\phi$ states is a robust, reproducible property of the network topology rather than an artifact of a single initial state. Crucially, the ablation experiment ($I_{tonic}$ = 0.0 pA) provides causal evidence that external sensory input alone is insufficient to maintain baseline conscious capacity; continuous endogenous neuromodulation (simulated via $I_{tonic}$ = 7.0 pA) is required to keep the Default Mode Network in a "primed" state capable of integrating sensory perturbations and recovering autonomous dynamics post-stimulus.

A critical theoretical limitation must be explicitly acknowledged: our model represents a functional simulation of conscious dynamics, not an actual conscious system. In philosophical terms, the execution of this 10,000-neuron network on digital hardware constitutes a computational artifact—or a functional "philosophical zombie" (P-zombie) at the micro-scale (Chalmers, 1996).While the continuous variable $\phi(t)$ quantifies the mathematical informational capacity and covariance structure of the network, the underlying physical substrate (silicon-based semiconductor logic gates) experiences no subjective, first-person phenomenology. The system exhibits access consciousness—the algorithmic manipulation and integration of structural states—without possessing phenomenal consciousness (the subjective "what-it-is-like" to experience a Qualia state) (Block, 1995; Chalmers, 1996).

Our simulation results intersect directly with the core philosophical debate surrounding artificial consciousness, highlighting the tension between functionalism/dualism and biological naturalism.

In agreement with Chalmers (1996) consciousness is a fundamental property of the universe that supervenes on functional organization rather than specific physical material (the principle of organizational invariance). From this prospective, if a computational network faithfully replicates the functional causal topology and integrated information ($\phi$) of a biological brain, it should, in principle, instantiate phenomenal experience. Under Chalmers' framework, our mathematical formulation of $\phi$ serves as a structural metric for the "Hard Problem" (Chalmers, 1995). However, because digital computers process algorithms sequentially via classical hardware, our numerical implementation remains a structural copy—a P-zombie whose non-zero $\phi$ values represent an informational proxy rather than subjective feel.

In contrast, John Searle (1980, 1992) contends that simulation is not duplication. Analogous to his famous Chinese Room thought experiment, a computer executing differential equations $({}^{dv}\!/_{dt}, {}^{du}\!/_{dt})$ and matrix operations, i.e. $\det(\sum)$, merely manipulates uninterpreted syntax without semantic understanding or subjective awareness. For Searle, consciousness is an exclusively biological phenomenon produced by specific causal power inherent to biological neural substrates (such as membrane biochemistry and quantum/molecular interactions). Under Searle's framework, calculating $\phi$= 15 bits on a digital computer is no more conscious than a computer simulation of a rainstorm is wet.

By evaluating $\phi$ within an algorithmic framework, our model demonstrates how complex information integration can be generated synthetically without resolving the substrate dependency of phenomenal experience.

To advance this elementary framework toward more realistic models of brain dynamics, several key extensions are planned for future work. To bridge the gap between software simulation and physical implementation, future iterations will port the spiking DMN architecture onto neuromorphic hardware chips (e.g., Intel Loihi or BrainScaleS). The reason for this is simple: Neuromorphic hardware operates with real physical voltage dynamics, continuous time, and massive parallelism, offering a physical substrate closer to biological computation (Indiveri et al., 2011).

Furthermore, incorporating Spike-Timing-Dependent Plasticity (STDP) will allow synaptic weights ($S_{ij}$) to evolve dynamically based on co-activation history. This will enable the network to form long-term memory traces and self-organize its top-down modulatory weights over prolonged sensory exposure.

Building upon these theoretical and hardware advances, a critical biological avenue is connecting this computational model to closed-loop bio-hybrid platforms (e.g., 3D cortical organoids integrated with microelectrode arrays). Comparing the simulated trajectories of $\phi(t)$ against empirical electrophysiological recordings from living human organoid networks will provide crucial biological validation of our DMN-sensory integration hypothesis (Smirnova et al., 2023).

## Code Availability

The source code and simulation models generated during the current study are publicly available in the GitHub repository: https://github.com/ResearchCodesHub/SNN-DMN-Qualia-Model

## Acknowledgments

I would like to express my sincere gratitude to Google's Gemini AI for its invaluable collaboration throughout this research. Gemini played a key role in developing the Python implementation for the spiking neural network (SNN) model simulating self-consciousness, the Default Mode Network (DMN), and the Integrated Information Theory (Φ) proxy for Qualia. Additionally, its assistance in generating structural diagrams and interactive visual simulations significantly contributed to the conceptual clarity and technical depth of this work.